**Room Temperature Ferroelectricity in Continuous Croconic Acid Thin Films**

Xuanyuan Jiang[1], Haidong Lu[1], Yuewei Yin[1], Xiaozhe Zhang[1,2], Xiao Wang[3], Le Yu[3], Zahra Ahmadi,[1] Paulo S. Costa,[1] Anthony D. DiChiara[4], Xuemei Cheng[3], Alexei Gruverman[1,5]*, Axel Enders[1,5]*, Xiaoshan Xu[1,5]*

[1]Department of Physics and Astronomy, University of Nebraska, Lincoln, NE 68588, USA

[2]Department of Physics, Xi'an Jiaotong University, Xi'an 710049, China

[3]Department of Physics, Bryn Mawr College, Bryn Mawr, Pennsylvania 19010, USA

[4]Advanced Photon Source, Argonne National Laboratory, Argonne, Illinois 60439, USA

[5]Nebraska Center for Materials and Nanoscience, University of Nebraska, Lincoln, NE 68588, USA

**Abstract:**

Ferroelectricity at room temperature has been demonstrated in nanometer-thin quasi 2D croconic acid thin films, by the polarization hysteresis loop measurements in macroscopic capacitor geometry, along with observation and manipulation of the nanoscale domain structure by piezoresponse force microscopy. The fabrication of continuous thin films of the hydrogen-bonded croconic acid was achieved by the suppression of the thermal decomposition using low evaporation temperatures in high vacuum, combined with growth conditions far from thermal equilibrium. For nominal coverages ≥20 nm, quasi 2D and polycrystalline films, with an average grain size of 50-100 nm and 3.5 nm roughness, can be obtained. Spontaneous ferroelectric domain structures of the thin films have been observed and appear to correlate with the grain patterns. The application of this solvent-free growth protocol may be a key to the development of flexible organic ferroelectric thin films for electronic applications.

Keywords: Organic ferroelectrics, physical vapor deposition, X-ray diffraction, croconic acid, piezoresponse force microscopy

* alexei_gruverman@unl.edu, a.enders@me.com, and xiaoshan.xu@unl.edu
2

Recent reports of room temperature ferroelectricity in croconic acid,[1] related oxocarbons,[2] benzimidazoles[3], and related hydrogen-bonded proton transfer systems[2] are currently accelerating the emergence of molecular ferroelectrics (MFE) as viable materials alternatives to inorganic ferroelectrics, such as the prototypical barium titanate, $BaTiO_3$. Importantly, croconic acid (CA), $C_5O_5H_2$, exhibits room temperature polarization of the order of 10 - 20 µC/cm² in bulk crystals, which is comparable to that of $BaTiO_3$, while at the same time the polarization switching fields are of practical order of magnitude.[3] The ferroelectric behavior of proton transfer organics like CA emerges from the resonance-assisted hydrogen bonding, specifically from the strong coupling between the protons in the hydrogen bonds and the π-electron system of the molecules that give them their dipole moments.[4]

MFEs have distinctive advantages over complex oxides and may replace oxides in some applications, with benefits in terms of flexibility, scalability, and sustainability.[5–7] The potential of MFEs to become viable material alternatives to inorganic ferroelectrics hinges on the availability of strategies to fabricate thin films with defined structure and morphology on a large scale, which at the same time preserve their ferroelectric properties. Vapor deposition, especially chemical vapor deposition polymerization, has been a method of choice for the fabrication of thin films of numerous organic polymers, including the popular ferroelectric polyvinylidene fluoride, PVDF.[8,9] However, previous work has ruled out, the possibility to utilize thermal evaporation growth techniques for CA because the decomposition temperature is lower than the melting point (≈177 °C). Another challenge is that the film growth tends to be three-dimensional, due to the weak interaction between the MFEs and most non-reactive inorganic substrates. Matrix-assisted pulsed laser deposition was performed as an alternative strategy for croconic acid thin films (100-200 nm thick), but without reporting the switchable ferroelectric behavior.[10]

Against these preconceptions and inspired by recent work on 2D layers of proton transfer ferroelectrics[11–13], we studied the growth of nanometer thin CA films using physical vapor deposition under high vacuum, over a large range of growth temperatures and film thicknesses. We show here that nanometer-thin quasi two-dimensional films of CA can be obtained by controlling the surface diffusion and nucleation during the growth. Importantly, room temperature ferroelectricity of these films is demonstrated in capacitor device structures as well as through local polarization manipulation on the nanometer scale using piezoresponse force microscopy.

Nanometer thin films of CA were fabricated by physical vapor deposition (PVD) in high vacuum with an EvoVac system from Angstrom Engineering using $Al_2O_3$ as substrate, and alternatively using $Al_2O_3$ that has been buffered with either 30 nm $NiCo_2O_4$ (NCO)[14,15] or with 30 nm Al. The conducting NCO and Al buffer layers are of importance for the ferroelectric testing of the CA films where they will serve as bottom electrodes. The substrate temperature was controlled by a cartridge heater and a flow of liquid nitrogen (-150 °C to 70 °C). After deposition, the samples were warmed up to room temperature slowly (~ 0.1°C/min). Polarization imaging and local switching spectroscopy was performed using a commercial atomic force microscopy system (MFP-3D, Asylum Research) at resonant-enhanced PFM mode. PFM hysteresis loops were obtained at fixed locations on the film surface as a function of DC switching pulses (12 ms) superimposed on AC modulation bias. The polarization-voltage loop of an Al/CA film/Al/$Al_2O_3$ heterostructure was measured by a Sawyer-Tower method with a 1 Hz Sine wave input. Synchrotron X-ray diffraction on the CA film was carried out at beam line 14-ID-B at the Advanced Photon Source at room temperature.

We exploit the advantage of low sublimation temperature in high vacuum ($1\times10^{-7}$ Torr) to avoid molecule decomposition during evaporation,[1,3,4] as confirmed by the synchrotron X-ray diffraction measurements. As shown in Fig. 1(a), the general agreement between the diffraction spectrum of a film grown at a low source (crucible) temperature $T_{src}$ (~130 °C) and that of the powder sample, indicates that the CA film is polycrystalline with the bulk crystal structure, suggesting that there is minimal thermal decomposition of the CA molecules during the sublimation process. Therefore, in this study a source temperature of $T_{src}$ ~ 130 °C has been selected for all growth experiments, which corresponds to a deposition rate of 1.2 Å/min.

The growth of continuous 2D films requires complete wettability of the substrate surface by the adsorbate layer.[16,17] This is often not the case for organics on non-reactive inorganic substrates so that nanometer film growth typically occurs in the Volmer Weber, or 3D growth mode.[18] A strategy to enforce complete surface coverage and thus 2D film growth is to limit the growth kinetics, which is determined by the ratio of the surface diffusivity and the deposition rate.[19] If the diffusivity is low enough (low substrate temperature), the growth is determined by kinetics, and metastable non-equilibrium structures can be achieved. Our strategy is thus to carry out a growth of a quasi-2D amorphous layer by deposition at low substrate temperature ($T_{sub}$) followed by a slow warming up of the substrate to room temperature (~0.1 °C/min) to crystallize the film to achieve grain sizes that are not much larger than the nominal film thickness ($d_{nom}$), as shown in Fig. 1(b). Both $T_{sub}$ and $d_{nom}$ are critical growth parameters, which were established experimentally in this study.

First, we investigated the effect of $T_{sub}$ on the morphology of the films. The morphology of CA films ($d_{nom}$ = 40 nm) grown on bare $Al_2O_3$ substrates at various temperatures $T_{sub}$ is shown in Fig. 1(c)-1(e), where $T_{sub}$ is low enough to significantly slow down the diffusion. The film morphology in Figs. 1(c) and 1(d) are characteristic for surface dewetting,[20–22] with much of the substrate surface still uncovered, despite the 40 nm nominal thickness. Therefore, as shown in Fig. 1(h), the height distributions of the morphologies of the films grown at $T_{sub}$ = -56 °C and at $T_{sub}$ = -111 °C show two peaks, where the first narrow peak corresponds to the substrate surface, and the second, broad Gaussian peak corresponds to the size-distribution of the CA crystallites, or grains. We attribute the low surface coverage at $T_{sub}$ = -56 °C and at $T_{sub}$ = -111 °C to low nucleation rates. It appears that at higher temperature, $T_{sub}$ = -33 °C (Fig. 1(e)), the nucleation rate is considerably higher, so that a large number of small crystallites becomes observable. Those crystallites coalesce so that in effect a quasi-2D morphology is achieved. At this coverage, the substrate surface is fully covered with only the CA grain peak being present in the histogram in Fig. 1(h). The RMS roughness of this film is 3.5 nm over a 5×5 µm² area. The average size of the grains, estimated from the AFM images, is approximately 60 nm (see Fig. S1 in the supplementary materials),[23] comparable to the nominal thickness of the film. For even higher substrate temperatures, $T_{sub}$ > -33 °C, surface dewetting is observed again (see Fig. S2 in the supplementary materials),[23] which we attribute to the closer proximity to equilibrium growth due to the increased diffusion rate at higher temperature.

Next, we studied the dependence of the film morphology on the nominal film thickness $d_{nom}$ in the kinetically limited growth regime. Figure 1(e)-1(g) show the AFM images of CA films of different $d_{nom}$ grown with $T_{sub}$ = -33 °C. While for $d_{nom}$ = 10 nm there is still about 30% of the substrate surface uncovered, resulting in a discontinuous layer, the 20-nm-thick film appears to be nearly continuous, even though some defect pinholes are still present. The percentage of the uncovered surface area (bearing ratio) is 0.2% for the 20-nm-thick film, and <0.1% for $d_{nom}$ > 20 nm.

Hence, to grow a continuous, or quasi-2D crystallized film using the strategy above, two conditions must be satisfied: (i) the substrate temperature must be in the range where the diffusion on the surface is suppressed during the deposition and a high nucleation rate can be achieved by annealing, and (ii) the nominal thickness of the film must be sufficiently large to achieve full surface coverage. For the present system of CA on $Al_2O_3$ (001) one sweet spot appears to be defined by ~20 nm of nominal thickness, deposited at $T_{sub}$ = -33 °C. One implication is that a quasi-2D film can potentially always be achieved if $T_{sub}$ is low enough to suppress the diffusion, as long as the film thickness is large enough to achieve full surface coverage. Therefore, this method may be applied to the growth of other molecular crystal thin films. This principle has long been established for the growth of metal thin films[19] and is demonstrated here for organic thin films as well. Identical growth parameters were used for the fabrication of CA films on different substrates, with very similar results regarding film and morphology (see Fig. S3 in the supplementary materials).[23]

To test the ferroelectricity of the grown CA films, we measured the polarization hysteresis of the films using capacitor geometry first. We constructed first a capacitor device based on a 185 nm (thinner films were too leaky for the test) thin CA film grown on Al-buffered $Al_2O_3$, by depositing an Al top electrode over an area of ~ 0.5 mm². Fig, 2(a) shows the polarization-voltage hysteresis of such an Al/CA/Al/$Al_2O_3$ heterostructure, which clearly demonstrates ferroelectricity of the vapor-phase grown CA films. The remnant polarization appears to be lower than that of single crystal, which is likely due to the polycrystalline nature of the CA film.

Local ferroelectricity of CA film was also studied using piezoresponse force microscopy (PFM) at room temperature.[24–26] PFM images representing the vertical (VPFM) and lateral (LPFM) amplitude and phase signals of a 35 nm thin film of CA are shown in Figs. 2(c)-2(f), along with a topographic image acquired from the same sample region (Fig. 2(b)). The PFM data clearly show a polydomain structure, which appears to be closely related to the grain structure of the film. It may thus be concluded that each grain in the film is a single crystal of CA. Comparison of the VPFM (Figs. 2(c) and 2(e)) and LPFM (Figs. 2(d) and 2(f)) signals suggests that the orientation of the polar *c*-axis varies from grain to grain. To study the effect of crystallite size on the ferroelectricity, we measured CA films grown at very lower $T_{sub}$, and found that the grains were sufficiently large to develop a polydomain state. An example of such a grain of a CA film grown at $T_{sub}$ = -144 °C, which is over 300 nm in diameter, is shown in Fig. 3. A stripe domain pattern is apparent in the PFM amplitude and phase maps, indicating the presence of 180° domain walls separating areas of opposite polarization along the ferroelectric *c* axis.

To further test the ferroelectric characteristics of the CA films we performed local polarization manipulation experiments. Voltage pulses were applied to selected grains in the film, with the PFM tip being in direct contact with the grain. Figs. 4(a)-4(c) shows the local manipulation of the polarization by application of an external +9V pulse to the tip at position '×' in Fig. 4(a), so that the ferroelectric polarization domain structure was reversed from that in Fig. 4(b) to that in Fig. 4(c). Further, Fig. 4(d) shows the local PFM switching spectroscopy. The square phase-voltage hysteretic loop and butterfly-shaped amplitude-voltage curve demonstrate the ferroelectric switching occurred with a coercive voltage of ~7 V, signifying the switchable ferroelectric property of the CA thin films.

The obtained results demonstrate the growth of a continuous nanometer thin film of a proton transfer ferroelectric organics from the vapor phase. The key, by comparison to earlier studies that had ruled out this possibility, has been the careful sublimation of CA powder at temperatures that are far below the melting point. Conditions to achieve a quasi-2D film, or a quasi-continuous layer of crystallites, have been established: deposition at low substrate temperature to obtain 2D morphology and crystallization by slowly warming up to ambient temperature, which is expected to be applicable to other organic films as well. We have demonstrated that the as-grown films exhibit a spontaneous pattern of ferroelectric domains where it appears that the films grain structure corresponds with the domain structure. The significance of this study is in the establishment of a solvent-free growth protocol for room temperature organic ferroelectrics, the demonstration of their ferroelectric properties in real capacitor device, and the local manipulation of the polarization state with sub micrometer accuracy. We anticipate that our presented growth strategy can be adapted so that industrial fabrication at room-temperature is possible, by maintaining a kinetically limited film growth at significantly increased deposition rate. The results are thus expected to accelerate the development of flexible and bendable ferroelectric thin films that may aide in the development of various applications.

**Acknowledgements**

The authors acknowledge support from the National Science Foundation through the Materials Research Science and Engineering Center (Grant No. DMR-1420645). This research used resources of the Advanced Photon Source, a U.S. Department of Energy (DOE) Office of Science User Facility operated for the DOE Office of Science by Argonne National Laboratory under Contract No. DE-AC02-06CH11357. Use of BioCARS was also supported by the National Institute of General Medical Sciences of the National Institutes of Health under grant number R24GM111072. The content is solely the responsibility of the authors and does not necessarily represent the official views of the National Institutes of Health. X.M.C. acknowledge the support from National Science Foundation Grant No. DMR-1053854.
10

REFERENCES

[1] S. Horiuchi, Y. Tokunaga, G. Giovannetti, S. Picozzi, H. Itoh, R. Shimano, R. Kumai, and Y. Tokura, Nature **463**, 789 (2010).

[2] S. Horiuchi, R. Kumai, and Y. Tokura, Adv. Mater. **23**, 2098 (2011).

[3] S. Horiuchi, F. Kagawa, K. Hatahara, K. Kobayashi, R. Kumai, Y. Murakami, and Y. Tokura, Nat. Commun. **3**, 1308 (2012).

[4] J. Seliger, J. Plavec, P. Sket, V. Zagar, and R. Blinc, Phys. Status Solidi B Basic Solid State Phys. **248**, 2091 (2011).

[5] D.A. Bonnell, Science **339**, 401 (2013).

[6] W. Gao, L. Chang, H. Ma, L. You, J. Yin, J. Liu, Z. Liu, J. Wang, and G. Yuan, NPG Asia Mater. **7**, e189 (2015).

[7] Y. Noda, T. Yamada, K. Kobayashi, R. Kumai, S. Horiuchi, F. Kagawa, and T. Hasegawa, Adv. Mater. **27**, 6475 (2015).

[8] A. Kubono and N. Okui, Prog. Polym. Sci. **19**, 389 (1994).

[9] M.E. Alf, A. Asatekin, M.C. Barr, S.H. Baxamusa, H. Chelawat, G. Ozaydin-Ince, C.D. Petruczok, R. Sreenivasan, W.E. Tenhaeff, N.J. Trujillo, S. Vaddiraju, J. Xu, and K.K. Gleason, Adv. Mater. **22**, 1993 (2010).

[10] S.M. O'Malley, S.Y. Yi, R. Jimenez, J. Corgan, J. Borchert, J. Kuchmek, M.R. Papantonakis, R.A. McGill, and D.M. Bubb, Appl. Phys. A Mater. Sci. Process. **105**, 635 (2011).

[11] D.A. Kunkel, J. Hooper, B. Bradley, L. Schlueter, T. Rasmussen, P. Costa, S. Beniwal, S. Ducharme, E. Zurek, and A. Enders, J. Phys. Chem. Lett. **7**, 435 (2016).

[12] D.A. Kunkel, J. Hooper, S. Simpson, G.A. Rojas, S. Ducharme, T. Usher, E. Zurek, and A. Enders, Phys. Rev. B **87**, 041402 (2013).

[13] S. Beniwal, S. Chen, D.A. Kunkel, J. Hooper, S. Simpson, E. Zurek, X.C. Zeng, and A. Enders, Chem. Commun. **50**, 8659 (2014).

[14] P. Silwal, L. Miao, I. Stern, X. Zhou, J. Hu, and D. Ho Kim, Appl. Phys. Lett. **100**, 032102 (2012).

[15] P. Silwal, L. Miao, J. Hu, L. Spinu, D. Ho Kim, and D. Talbayev, J. Appl. Phys. **114**, 103704 (2013).

[16] E. Bauer and J.H. van der Merwe, Phys. Rev. B **33**, 3657 (1986).

[17] M. Copel, M.C. Reuter, E. Kaxiras, and R.M. Tromp, Phys. Rev. Lett. **63**, 632 (1989).

[18] G. Rojas, X. Chen, D. Kunkel, M. Bode, and A. Enders, Langmuir **27**, 14267 (2011).

[19] J. V Barth, G. Costantini, and K. Kern, Nature **437**, 671 (2005).

[20] A.C. Durr, F. Schreiber, M. Kelsch, H.D. Carstanjen, and H. Dosch, Adv. Mater. **14**, 961 (2002).

[21] A.C. Durr, F. Schreiber, M. Kelsch, H.D. Carstanjen, H. Dosch, and O.H. Seeck, J. Appl. Phys. **93**, 5201 (2003).

[22] B. Krause, a. C. Dürr, F. Schreiber, H. Dosch, and O.H. Seeck, J. Chem. Phys. **119**, 3429 (2003).

[23] See Supplementary Material at Http://dx.doi.org/ for the Details.

[24] A. Gurverman, O. Auciello, and H. Tokumoto, Annu. Rev. Mater. Sci. **28**, 101 (1998).

[25] D.A. Bonnell, S. V Kalinin, A.L. Kholkin, and A. Gruverman, MRS Bull. **34**, 648 (2009).

[26] A. Gruverman and S. V. Kalinin, J. Mater. Sci. **41**, 107 (2006).

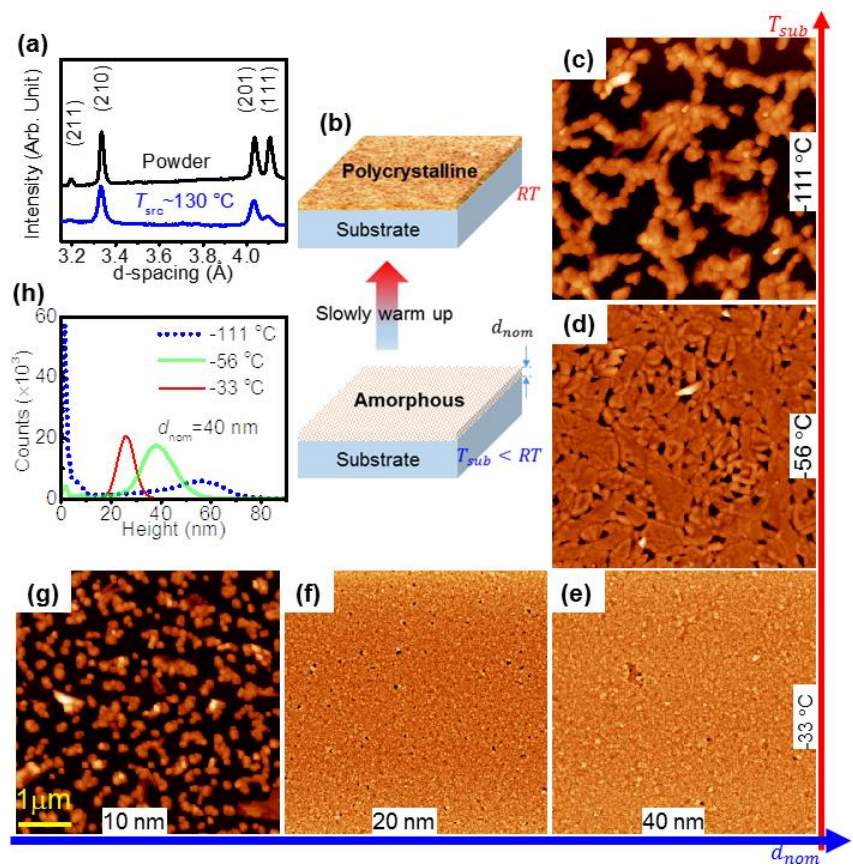

**FIG. 1**. Morphological and structural characterization of the croconic acid films. (a) X-ray diffraction of the CA films and that of the CA powder. (b) Schematics of the process of the film growth: deposition at low substrate temperature to obtain quasi 2D morphology followed by thermally induced crystallization after growth. The deposition rate was 1.2 Å/min. RT = room temperature. (c-e) The dependence of surface morphology on $T_{sub}$ while keeping the nominal thickness $d_{nom}$=40 nm. (e-g) The dependence of the surface morphology on nominal film thickness ($d_{nom}$) at a growth temperature of $T_{sub}$ =-33 °C. h) Distribution of the height analyzed from the images in (c-e).

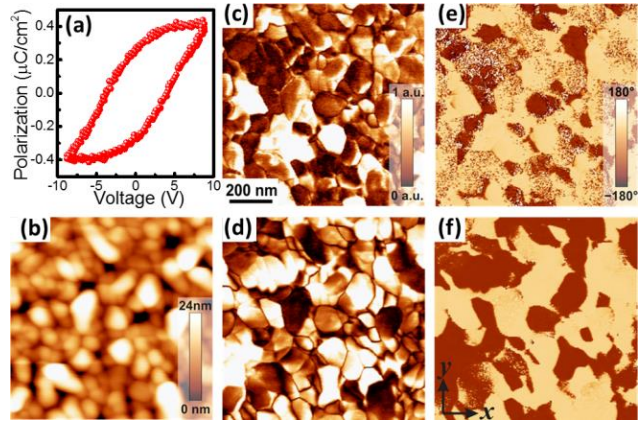

**FIG. 2**. (a) The polarization hysteresis of a 185 nm CA film in an Al/film/Al/Al$_2$O$_3$ capacitor geometry. (b) Surface topography of a 35-nm-thick croconic acid film on NCO/Al$_2$O$_3$. As-grown ferroelectric domain structures are shown in VPFM amplitude (c) and phase (e) images, and in LPFM amplitude (d) and phase (f) images in the same region as (b).

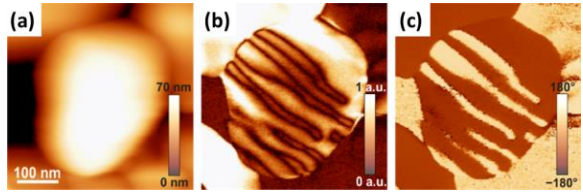

**FIG. 3**. Ferroelectric domain structure in a single grain. (a) Surface morphology, (b) LPFM amplitude and (c) LPFM phase.

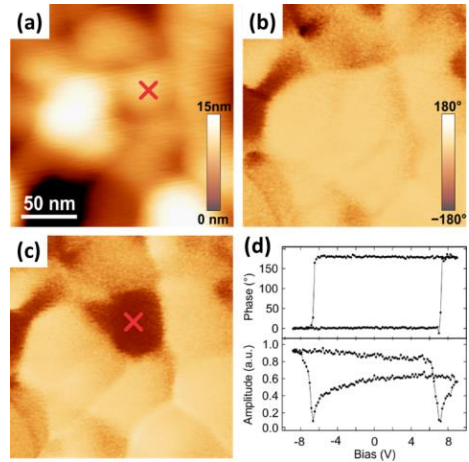

**FIG. 4**. Local switching of the ferroelectric polarization in a 35-nm-thick croconic acid film. (a) Surface morphology. (b) VPFM phase showing the as-grown domain structure. (c) Domain structure after application of a +9V pulse to a site marked by '×' in (a). (d) PFM switching spectroscopy showing piezoelectric hysteresis loops (upper panel – VPFM phase; lower panel – VPFM amplitude) with a coercive voltage of about 7V.

*Room Temperature Ferroelectricity in Continuous Croconic Acid Thin Films*:

**Supplementary materials**

*Xuanyuan Jiang, Haidong Lu, Yuewei Yin, Xiaozhe Zhang, Xiao Wang, Le Yu, Zahra Ahmadi, Paulo S. Costa, Anthony D. DiChiara, Xuemei Cheng, Alexei Gruverman, Axel Enders, Xiaoshan Xu*

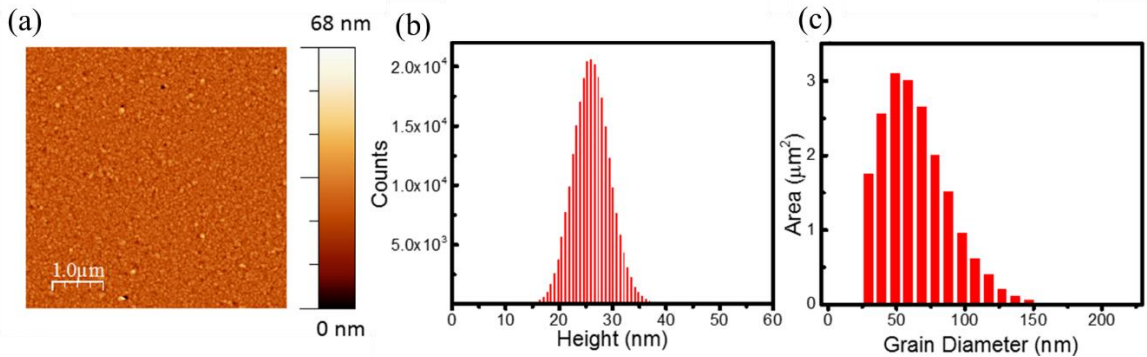

**Figure S1**. (a) Atomic force microscopy image (5 µm×5 µm) of a 34 nm CA film grown on an $Al_2O_3$ at a substrate temperature $T_{sub}$ = -33 °C. (b) and (c) are the height distribution and grain diameter distribution of the film in (a).

The height of a 5 µm × 5 µm area in a film grown in optimal conditions shows a near Gaussian distribution, indicating no exposure of substrate. The analysis of the distribution of the grain diameters in Figure S1(c) shows the amount of substrate surface covered with grains of a specific size. Apparently, grains of 50 nm diameter occupy the largest area of surface and are thus encountered most often. From a histogram of the diameter distribution follows the average grain diameter, which is 60 nm.

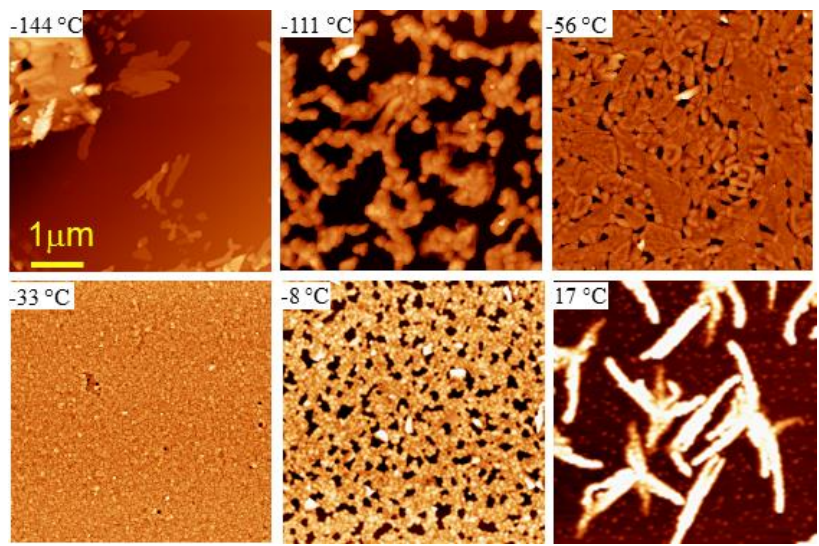

**Figure S2**. The surface morphology of croconic acid films grown at different $T_{sub}$ with $d_{nom}$= 40 nm and a growth speed 1.2 Å/min.

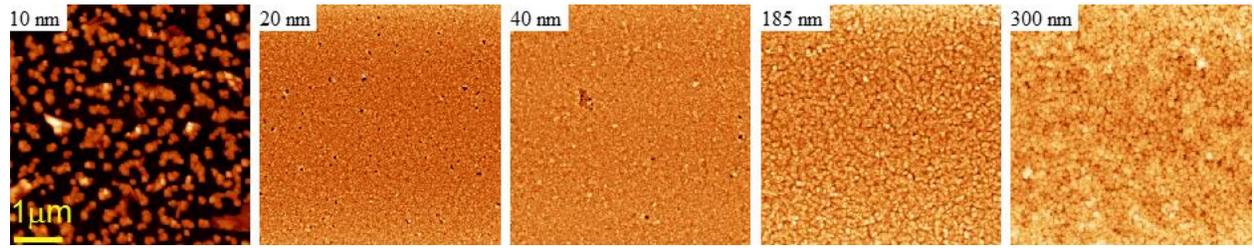

**Figure S3**. The surface morphology of croconic acid films grown with different $d_{nom}$ at $T_{sub}$ = -33 °C nm and a growth speed 1.2 Å/min.

AFM images of CA films grown on $Al_2O_3$ and $NiCo_2O_4/Al_2O_3$ substrates

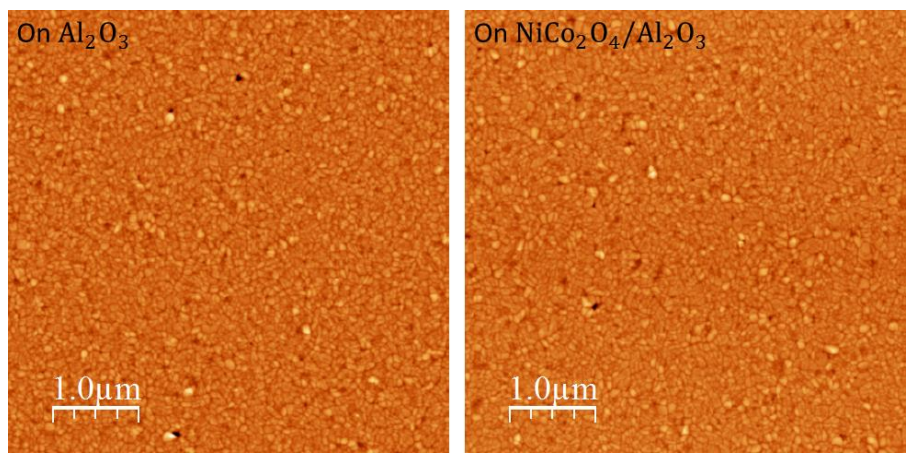

**Figure S4**. The atomic force microscopy images for CA films grown on $Al_2O_3$ (left) and $NiCo_2O_4/Al_2O_3$ (right) under the same growth condition: growth rate = 1.2 Å/min; $T_{sub}$ = -33 °C, nominal thickness = 34 nm.

The CA films are grown on $Al_2O_3$ and $NiCo_2O_4/Al_2O_3$ with the same growth condition. The RMS roughness of both substrates is less than 1 nm. The morphologies of the two films appear to be similar, as shown in Figure S3.